\documentclass[sigplan,screen]{acmart}
\usepackage{graphicx}
\usepackage{fontawesome5}
\usepackage{subfigure}

\setcopyright{none}
\renewcommand\footnotetextcopyrightpermission[1]{} %

\setcopyright{acmlicensed}\acmConference[ASPLOS '24 WACI]{29th ACM International Conference on Architectural Support for Programming Languages and Operating Systems, Volume 1}{April 27-May 1, 2024}{La Jolla, CA, USA}
\acmBooktitle{29th ACM International Conference on Architectural Support for Programming Languages and Operating Systems, Volume 1 (ASPLOS '24), April 27-May 1, 2024, La Jolla, CA, USA}

\begin{document}
\title{Apparate \faIcon{magic}: Evading Memory Hierarchy with GodSpeed Wireless-on-Chip}
\author{Nitesh Narayana GS}
\affiliation{
  \institution{\textit{Universitat Politècnica de Catalunya}}
  \country{Barcelona, Spain}
}
\email{nitesh@ac.upc.edu}

\author{Abhijit Das}
\affiliation{%
  \institution{\textit{Universitat Politècnica de Catalunya}}
  \country{Barcelona, Spain}
}
\email{abhijit.das@upc.edu}
\renewcommand{\shortauthors}{N. N. GS and A. Das}
\maketitle
\section{Introduction}
We have come a long way since the first digital computer ENIAC was developed in the early 1940s~\cite{eniac}. For example, our mobile phones have 100,000 times more computing power than the Apollo Guidance Computer (AGC)~\cite{apollo} that landed the first humans on the Moon~\cite{moon}. Over the past few decades, we have witnessed some remarkable technological and architectural advancements, from the inception of transistors to their continuous shrinking to date, from single-core to massive many-core processors, and from large monolithic chips to manageable chiplets. What is very intriguing is how these advancements often align with certain time-tested predictions, like Moore's Law~\cite{moore1965cramming}, Dennard Scaling~\cite{dennard1974design}, etc.

A current trend shows that the technological and architectural advancements have shifted the fundamental bottleneck of a system design from computation to communication~\cite{interconnect2023,interconnect2022,salahuddin2018era}. Chip and package-scale communication will soon start dictating the designs of next-generation computing systems. While everyone has put on their thinking caps to envision how future computing systems will look like, we present a wild and crazy yet calculated guess. We believe that by extrapolating the data from the time-tested predictions and current trends, we could predict or, even better, suggest how computing systems could be designed in 2050\footnote{By 2050, the transistor~\cite{transistor} would complete its 100$^{th}$ anniversary.}.

We present \textbf{Apparate}~\faIcon{magic}\footnote{Apparate or Apparition \cite{rowling2000harry} is an act of magical transportation from one place to another without any physical means in the Wizarding World~\cite{apparate}.}, a prediction-cum-concept to use wireless communication within the chip to evade memory hierarchy for superior performance and efficiency of computing systems. The subsequent sections will describe the motivation, feasibility, and implementation of Apparate.

\section{Trends Till Now to Trends Here After}
The projection of trends is restricted till 2050 to provide a comprehensive view of the extrapolations. This timeframe strikes a balance, offering a glimpse into the foreseeable future while maintaining anticipation and excitement.

\begin{figure}[t]
  \centering
  \includegraphics[width=0.8\columnwidth]{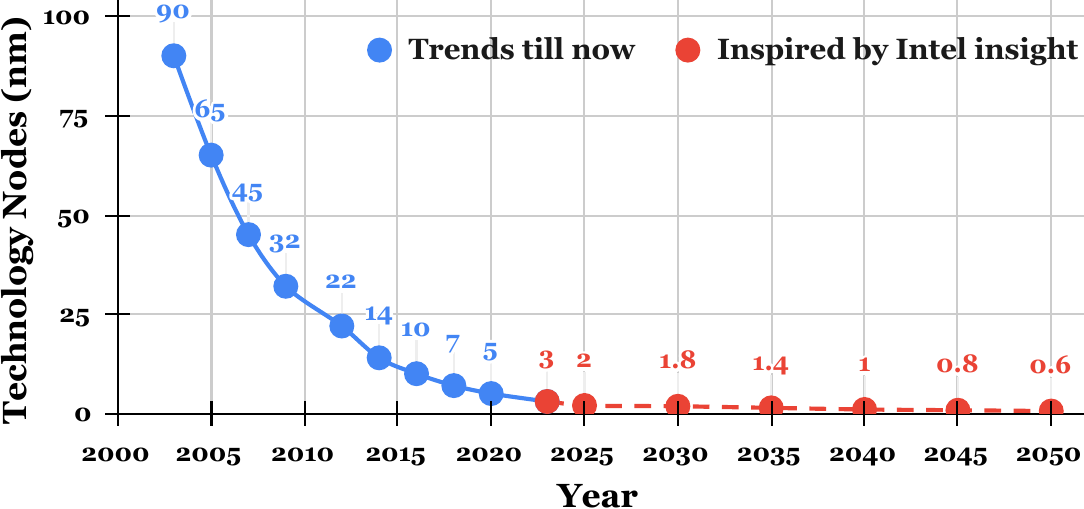}
  \vspace{-0.4cm}
  \caption{Yet another Moore's Law graph towards 2050.}
  \label{gra:moores}
\end{figure}

\begin{figure}[t]
  \centering
  \includegraphics[width=0.8\columnwidth]{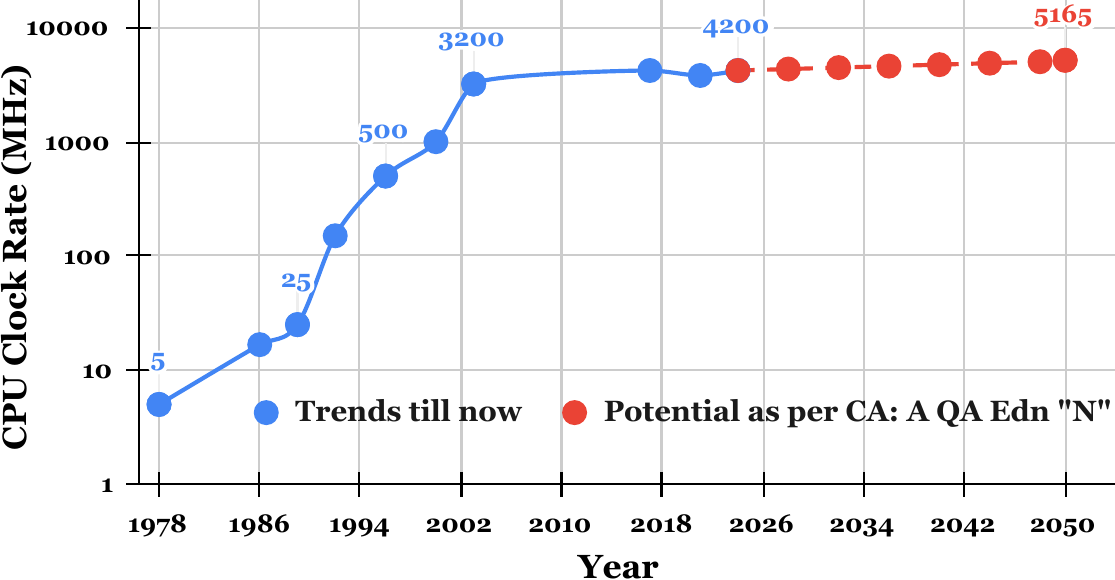}
  \vspace{-0.4cm}
  \caption{Processor clock rate towards 2050.}
  \label{gra:clock}
\end{figure}

\begin{figure}[t]
  \centering
  \includegraphics[width=0.8\columnwidth]{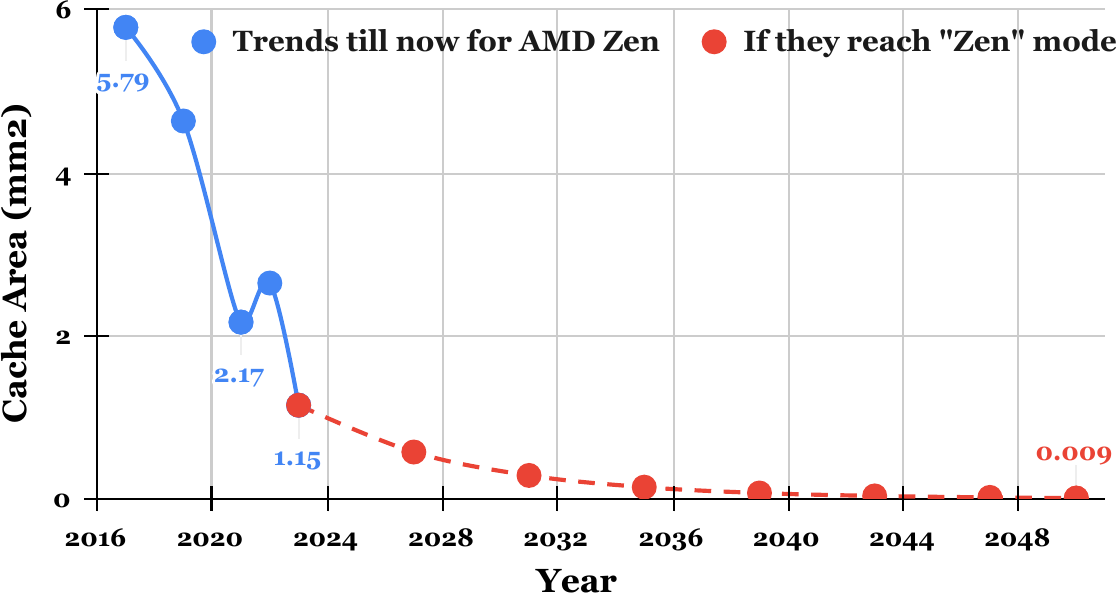}
  \vspace{-0.4cm}
  \caption{On-chip cache area towards 2050.}
  \label{gra:cache}
  \vspace{-0.5cm}
\end{figure}

\textbf{Transistor Scaling:} While most computer architects believe Moore's Law is either nearing or already dead, Intel~\cite{intel} disagrees. They believe that Moore's Law will continue to live on with the support of advanced packaging technology and new materials~\cite{moores}. We borrow their optimism and show in Figure~\ref{gra:moores} that Moore's Law might continue but at a slower pace as it will start to decelerate in 2025. Nevertheless, the extrapolated trajectory suggests that the number of transistors on a chip area will continue to increase. \textit{This trend is encouraging and will continue to allow computer architects to explore unconventional and ground-breaking chip design}.

\textbf{Processor Clock:} In Figure~\ref{gra:clock}, we have extrapolated Figure~1.11 of the book CA: A QA Edn ``6''~\cite{hennessy2017computer}, showing the growth in clock rate of microprocessors. We observe that after an exponential increase until the last decade, the processor clock rate has now become more or less stagnant. This is mainly due to the breakdown of Dennard Scaling. \textit{This trend implies that there isn't going to be much of a change in the required L1 cache bandwidth, which is currently at 1 Tbps}.

\textbf{On-Chip Cache Area:} To identify the trend in on-chip cache area footprint over the years, we examined the AMD Zen processor~\cite{zen} series. Extrapolating their L1+L2+L3 cache areas in Figure~\ref{gra:cache}, we observe a continued decrease in the footprint. While the L1 cache size remained steady to keep up with the processor speed, L2 and L3 kept increasing. Nevertheless, their overall footprint continued decreasing due to the exponential growth in transistor count within the same chip area. \textit{\underline{This trend raises an intriguing question}: With no area overhead in increasing cache size and a saturated L1 cache bandwidth, can DRAMs replace caches in the future?}

\begin{figure}[t]
  \centering
  \includegraphics[width=0.8\columnwidth]{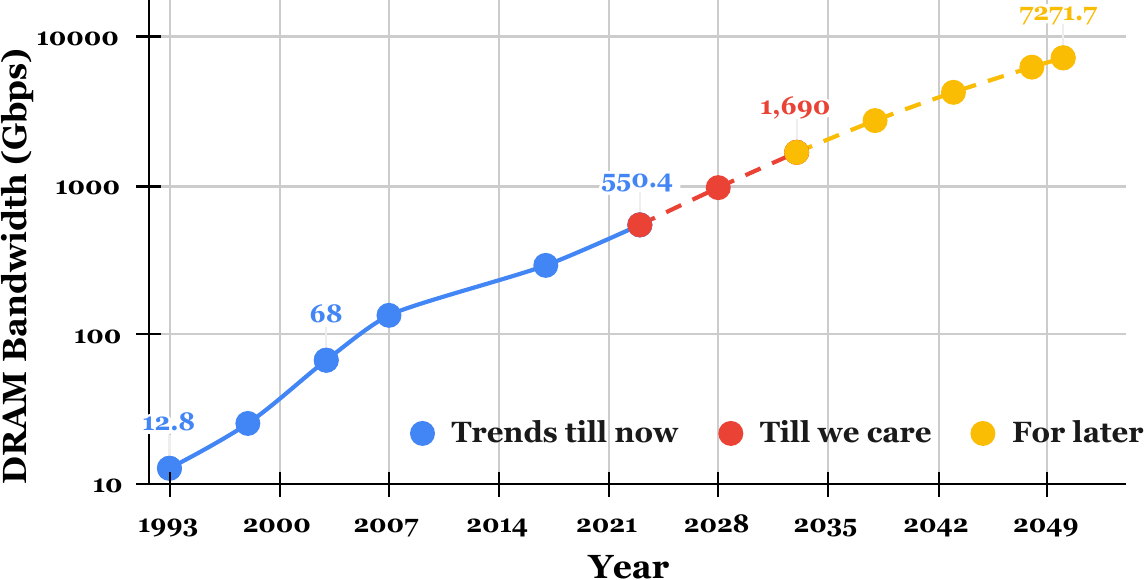}
  \vspace{-0.4cm}
  \caption{DDR ``N'' bandwidth towards 2050.}
  \label{gra:dram}
\end{figure}

\textbf{DRAM Bandwidth:} Our extrapolation in Figure~\ref{gra:dram} shows that DRAM could achieve 1 Tbps bandwidth well before 2050, with a projection of up to 7 Tbps by that time. A CPU directly communicating with DRAM would ideally remove all the memory hierarchy-related bottlenecks. However, despite this promising outlook, putting DRAMs alongside CPUs poses inherent challenges due to its capacitor-based storage mechanism. \textit{\underline{This roadblock raises a serious curiosity}: Could there be a way to make the CPU directly talk to the DRAM?}

\begin{figure}[t]
  \centering
  \includegraphics[width=0.8\columnwidth]{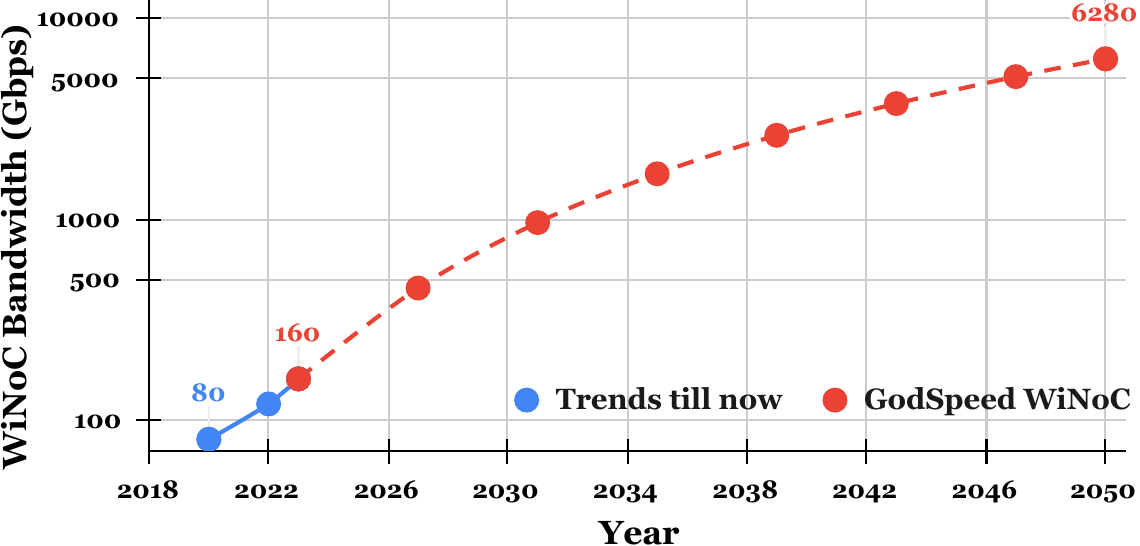}
  \vspace{-0.4cm}
  \caption{GodSpeed WiNoC bandwidth towards 2050.}
  \label{gra:winoc}
  \vspace{-0.5cm}
\end{figure}

\begin{figure}
  \centering
  \includegraphics[width=0.8\columnwidth]{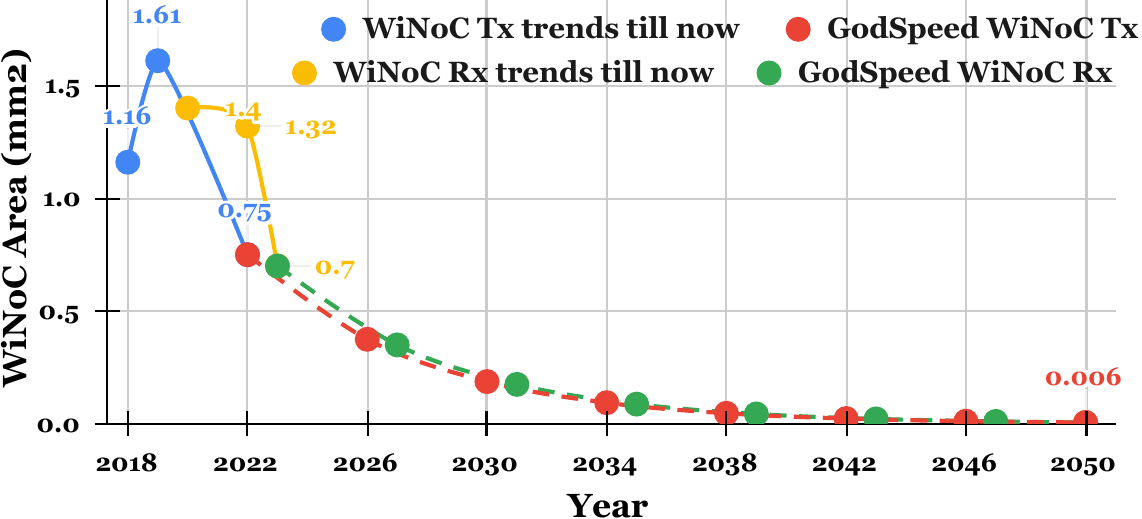}
  \vspace{-0.4cm}
  \caption{WiNoC Tx and Rx areas towards 2050.}
  \label{gra:trx}
  \vspace{-0.4cm}
\end{figure}

\textbf{Wireless Network-on-Chip:} There has been a growing interest in Wireless Network-on-Chip (WiNoC)~\cite{franques2021widir, fernando2019replica, abadal2016wisync, ganguly2010scalable}. To shed light on this emerging trend, we have charted the extrapolated trajectories of WiNoC's bandwidth and area in Figures~\ref{gra:winoc} and~\ref{gra:trx}, respectively. On one side, we observe that WiNoC could attain L1 cache-equivalent bandwidth of 1 Tbps as swiftly as DRAMs. On the other side, we also observe that the WiNoC Transmitter (Tx) and Receiver (Rx) could rival the cache area footprint by 2050. \textit{\underline{This prompts us to contemplate the unthinkable}: Could WiNoC replace caches and be the bridge between the CPU and DRAM?}

\begin{figure}
  \centering
  \includegraphics[width=0.5\columnwidth]{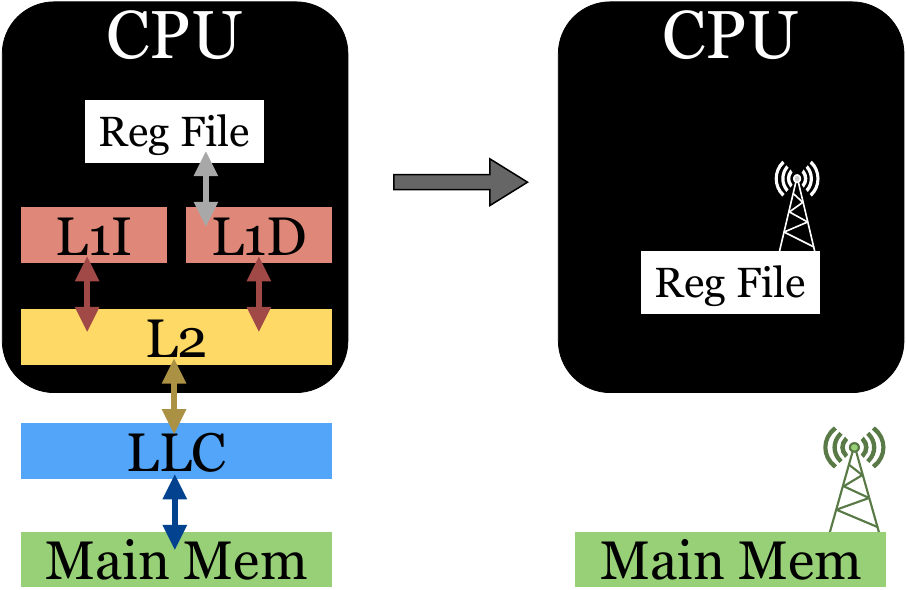}
  \vspace{-0.2cm}
  \caption{The Apparate concept.}
  \label{fig:apr}
  \vspace{-0.4cm}
\end{figure}
\section{Apparate: A Way Forward}
We propose a radical departure from traditional CPU architecture and advocate for the evasion of the ``\textit{latency-hiding}'' caches from the memory hierarchy. As illustrated in Figure~\ref{fig:apr}, our concept, \textbf{Apparate}~\faIcon{magic}, replaces caches with wireless transceivers at the register file and the DRAM. When the CPU initiates a memory request, it ``\textit{apparates}'' from the register file to the DRAM through the transceiver. The future high-bandwidth DDR ``N'' DRAM will immediately respond with the data, which will ``\textit{apparate}'' back to the register file. This arrangement will obviate the need for memory hierarchy\footnote{A long-standing dream of most computer architects!}, thereby removing its associated bottlenecks.

In terms of feasibility, Apparate prompts several intriguing questions. Below, we offer a teaser of potential questions and answers, leaving the wild exploration to the readers!
\begin{itemize}
    \item \textit{What happens to ``cache'' coherency?} As caches will be phased out, the onus for maintaining coherency will fall onto the main memory. Consequently, coherence tables will find their new home in the main memory.
    \item \textit{How hot can the chip now get?} Regardless, it should be more manageable with no caches to heat things up!
    \item \textit{Will we need prefetching, replacement, etc.?} We may no longer need conventional complexities like cache prefetching, replacement, etc. This will liberate valuable chip space, allowing for new innovations!
\end{itemize}
\section{Conclusion}
We hypothesise that computer architecture is poised to transition into an era dominated by WiNoC technology. Therefore, investing thought into Apparate will not only shape the future design of computing systems but also provoke a fundamental question: \textit{\underline{Do we truly need what we already have?}} As demonstrated through Apparate, it becomes evident that traditional caches\footnote{And hence as a butterfly effect other structures related to caches.} may no longer be essential in the future.
\begin{acks}
We extend our sincere gratitude to the WACI initiative of ASPLOS for giving us a forum that recognises thoughts and ideas that are indeed wild and crazy! Special thanks to our friend Hrishikesh R Menon~\cite{hrimenon} for his creative support.
\end{acks}
\bibliographystyle{ACM-Reference-Format}
\bibliography{references}
\end{document}